\documentclass{elsart}

 \usepackage{graphicx}

\usepackage{amssymb}
\usepackage{amsmath}
\usepackage{amsfonts}

\renewcommand\ln{\ell{n}}
\newcommand\beq{\begin{equation}}
\newcommand\eeq{\end{equation}}
\newcommand\bea{\begin{eqnarray}}
\newcommand\eea{\end{eqnarray}}
\newcommand\bseq{\begin{subequations}} 
\newcommand\eseq{\end{subequations}}
\newcommand\bal{\begin{align}}  
\newcommand\ealign{\end{align}}    

\renewcommand\ln{{\rm ln}}

\begin{document}

\begin{frontmatter}



\title{Complex dynamics of the biological rhythms: 
 gallbladder and heart cases}

\author[FII,ICRA]{Giovanni Imponente}%
	 \ead{imponente@icra.it}
%
%
%
\address[FII]{Dipartimento di Fisica, 
 						Universit\'a ``Federico II'', 
 						and INFN Napoli -- Italy}
\address[ICRA]{ICRA -- International Center for
							Relativistic Astrophysics, 
							c/o G9 - Dip. Fisica, 
							Universit\'a ``La Sapienza''
P.za A.Moro 5, I-00185 Roma}
%

\begin{abstract}

A theoretical analysis of the mechanisms 
underlying the dynamics of gallbladder 
and heart pulsation 
could clarify the question regarding 
the classification as chaotic of the 
associated behaviour, eventually 
related to a normal and 
healthy beat; this analysis 
is particularly relevant in view of 
the control of  dynamics bifurcations 
arising in situations of disease.
In this work is presented a summary of the 
DFA method applied to gallbladder volume 
data for a modest number of 
 healthy and ill patients: the presence of 
 signal correlation is found in both 
 cases, but the  fit shapes  
differ from some critical values.

\end{abstract}

\end{frontmatter}



Complex systems, as observed in nature 
and modelled by mechanical or electrical approximations, have a dynamics characterized 
 by dependence on many competing 
effects admitting multiple possible 
behaviours, so that the system 
tends to alternate among them.
The task of the description of such 
a variegate structure relies in  
the research of a manifest form of coherent 
structure, in order to recognize a hierarchy 
of patterns
over a wide range of time and/or length 
scales. 
Commonly comprehension of natural systems  
is based on highly simplified models testing  
linearised or tractable equations, though
the intrinsic approximations
of numerical methods to solve 
sets of differential equations
have to be taken with much care, 
because of  the unavoidable 
effect over the results due to the 
 high sensitivity to approximations 
 and to initial conditions of
chaotic systems.
Thus, a high level of 
unpredictability on the subsequent 
dynamics behaviour often remains,
especially when  studying a living system.\\
The goal in an  study 
of the data as well as of the model 
is to find a possible
clear distinction between the normal (healthy)
 and the pathological cases. \\
We consider the pulsation of the gallbladder
and of the cardiac muscle, with particular attention to the former as an example for 
application of the Detrended Fluctuation 
Analysis (DFA) method \cite{PHY} 
to data manipulation. 
For the latter, an analysis of the 
implications of fibres geometry over 
the cardiac muscle is needed, 
 in the framework of a theoretical model 
 for heart activity
 assuming several pacemakers 
bounded to the fibres in the
heart complex geometry. 
The methods so far 
adopted rely either on phenomenological 
models of waves propagating over
the cardiac muscle seen as a membrane 
\cite{BH96,H9798}, 
either on the application of data analysis 
methods \cite{Peng94} looking for 
traces of chaoticity or fractality. 
The idea of linking the healthy heart rate 
to a fractal behaviour is based on 
\cite{Wit95}, the ill dynamics
presenting a regular inter-beat interval, 
a re-entrant spiral wave is observable 
at the very last beat 
of the stopping the heart together 
with ventricular fibrillation \cite{Wiggers1930}.
Let us consider how to connect fractals to 
gallbladder in humans. The term \textit{fractal}
is associated to a geometrical object 
satisfying the criteria of \textit{self-similarity},
i.e. the existence of a sub-structure composed of 
sub-units resembling some statistical properties  
of the whole object, and 
\textit{fractal dimensionality}. 
The test to determine if a 2-dimensional curve 
is self-similar consists firstly of
taking a subset of the object and rescaling it 
to the same size of the original object, 
using the same magnification   	factor for both 
its width and height and, secondly, 
	comparing the statistical properties of the 
	rescaled object with the original object \cite{WTT}. 
	In contrast, to properly compare a 	subset 
	of a time series with the original data set, 
	we need \textit{two}
	 magnification factors (along the horizontal 
	 and vertical axes), 	since these two axes 
	 represent different physical variables. \\
	 A time-dependent process $y(t)$ (time series) 
	 is self-similar with self-similarity parameter 
	 $\alpha$  if
	 \beq
	 y(t) \equiv a^{\alpha} y\left(\frac{t}{a}\right) \, , 
	 \eeq 
i.e. $y(t)$ has the identical probability distribution 
as a properly rescaled process,  $a^{\alpha} y(t/a)$: 
a time series is rescaled on the $x$-axis by a factor 
$a$, $(t\rightarrow  t/a)$,  and on the $y$-axis by  
$ a^{\alpha}$,   $(y\rightarrow  a^{\alpha} y)$. \\
The method of DFA is based on this concept and 
permits to operate on ``real-world'' time series 
with a given procedure. 
For any given size of observation window, the time 
series is  divided into subsets of independent 
windows of the same size. 
 To obtain a more reliable estimation of the 
 characteristic fluctuation at this window size, 
 we average over all individual values of standard 
 deviations $s$ associated to different 
 sets of data and  obtained from these subsets, 
then we repeat these calculations 
for many different window sizes. 
The exponent   $\alpha$   is estimated by fitting a 
line on the log-log plot of $s$ versus 
the number of data $n$ belonging to each window  
across the relevant range of scales. 
We consider stationary time series, 
i.e. mean, standard deviation and higher moments, 
correlation functions are invariant under time translation.
The fractal analysis applied to physiologic time series is 
pursued for highly non-stationary  by an integration 
procedure which  will make the non-stationarity of 
the original data even more apparent.
The advantages of DFA over conventional methods 
(e.g., spectral analysis and Hurst analysis) 
rely on the detection of intrinsic correlation 
 embedded in a seemingly non-stationary time-series
 and on avoiding  spurious detection of apparent 
 correlation, which may be an artefact of extrinsic 
 trends \cite{Peng94}.
 The time-series of signals $B(i)$ is firstly integrated
and then the trend is computed via the function $F(n)$
\begin{align}
y(k)=\sum^k_{i=1}
[B(i)-B_{ave}]\, \qquad  
 F(n)= \sqrt{\frac{1}{N}\sum_{k=1}^{N}\left[y(k)-y_n(k)\right]^2}
\end{align}
which relates the least squares fluctuation in each box
of data to the number of points belonging to that box
and the correlation parameter is related to the 
mean square fluctuation as 
\beq
\alpha = \frac{\ln M_y}{\ln M_x}=
\frac{\ln s_2 -\ln s_1}{\ln n_2 -\ln n_1} \, ,
\eeq
for magnification windows along the $x$- and $y-$axis
$M_x$ and $M_y$, respectively, and standard deviations
$s_{1,2}$ corresponding to boxes of dimension 
$n_{1,2}$, and then is computable as the slope 
on a log-log plot of $F(n)$ versus $n$.
It is usually classified the level of correlation 
as: 
(i) $\alpha \in (0, 0.5)$ anticorrelation;
(ii.a) $\alpha \sim 0.5$ short-term 
exponential correlation; 
(ii.b) $\alpha = 0.5$
white noise; 
(iii) $\alpha \in ( 0.5,1)$ persistent 
long-range correlation; 
(iv.a) $\alpha >1$ correlation, 
not power law; 
(iv.b) Brown noise. 
\begin{figure}
\resizebox{\hsize}{0.15\vsize}{\includegraphics{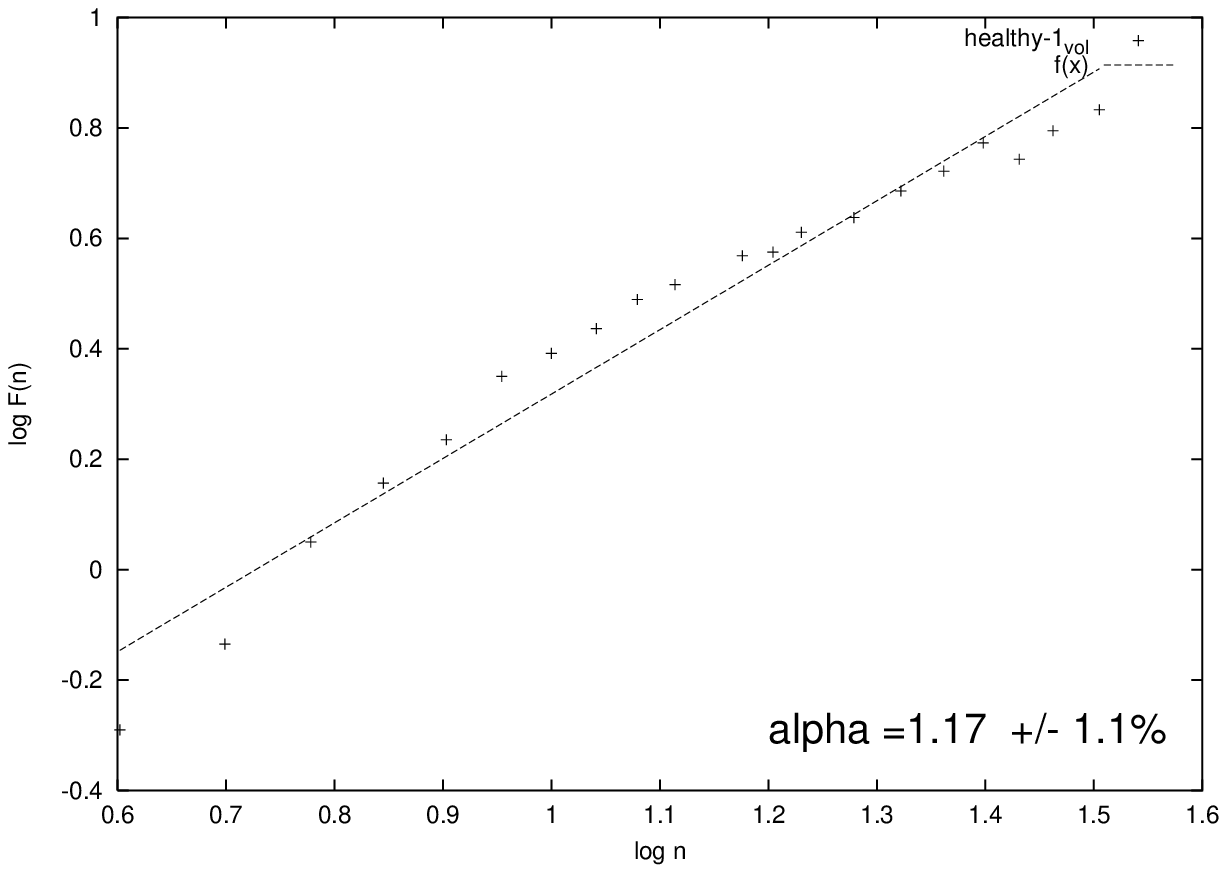}
									\includegraphics{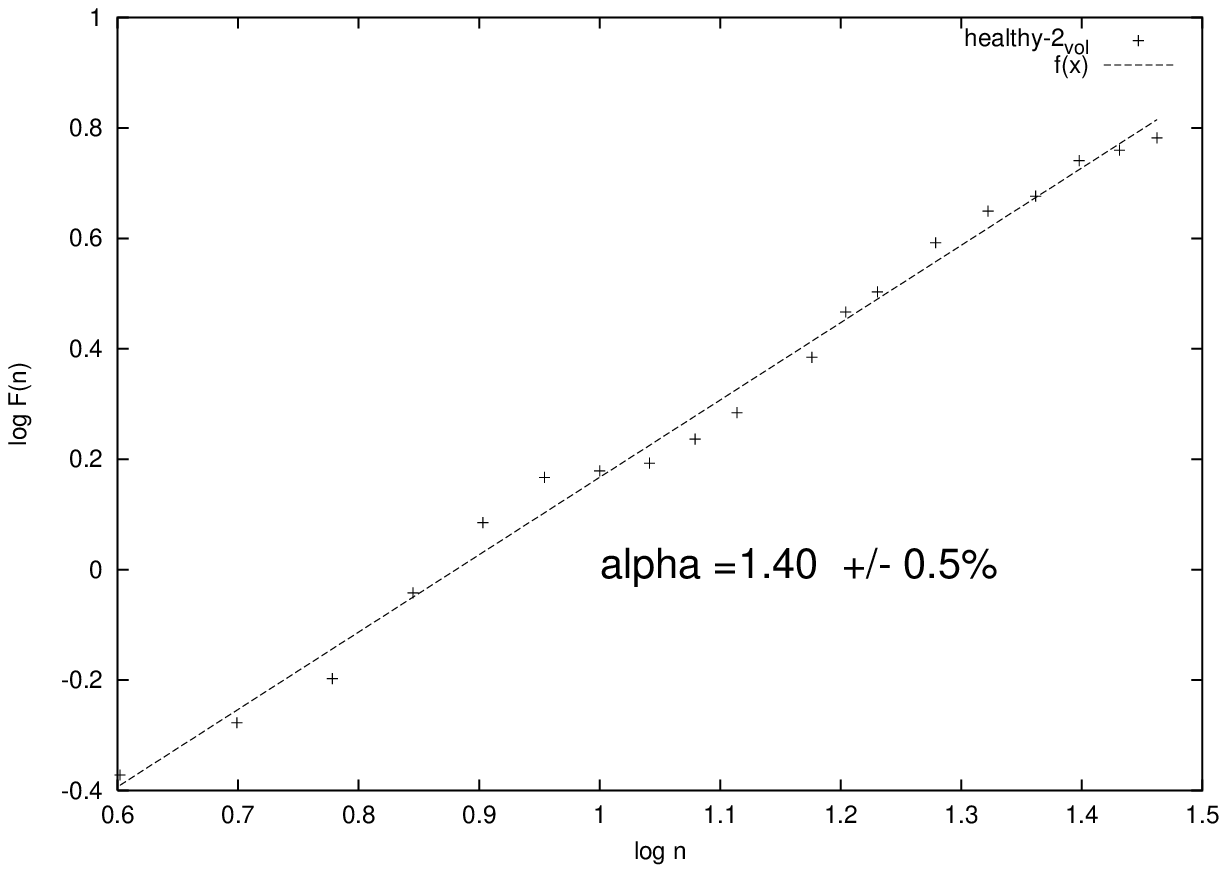}}
\resizebox{\hsize}{0.15\vsize}{\includegraphics{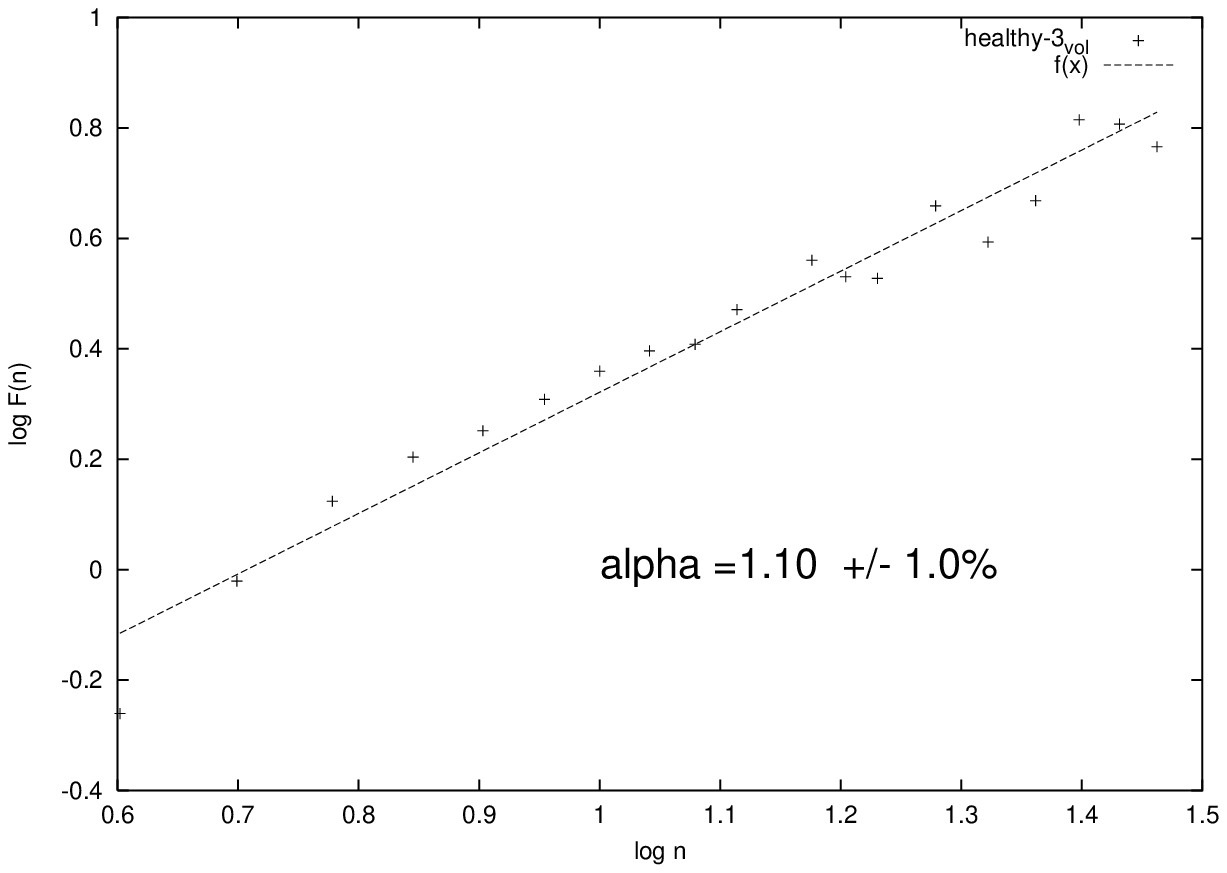}
									\includegraphics{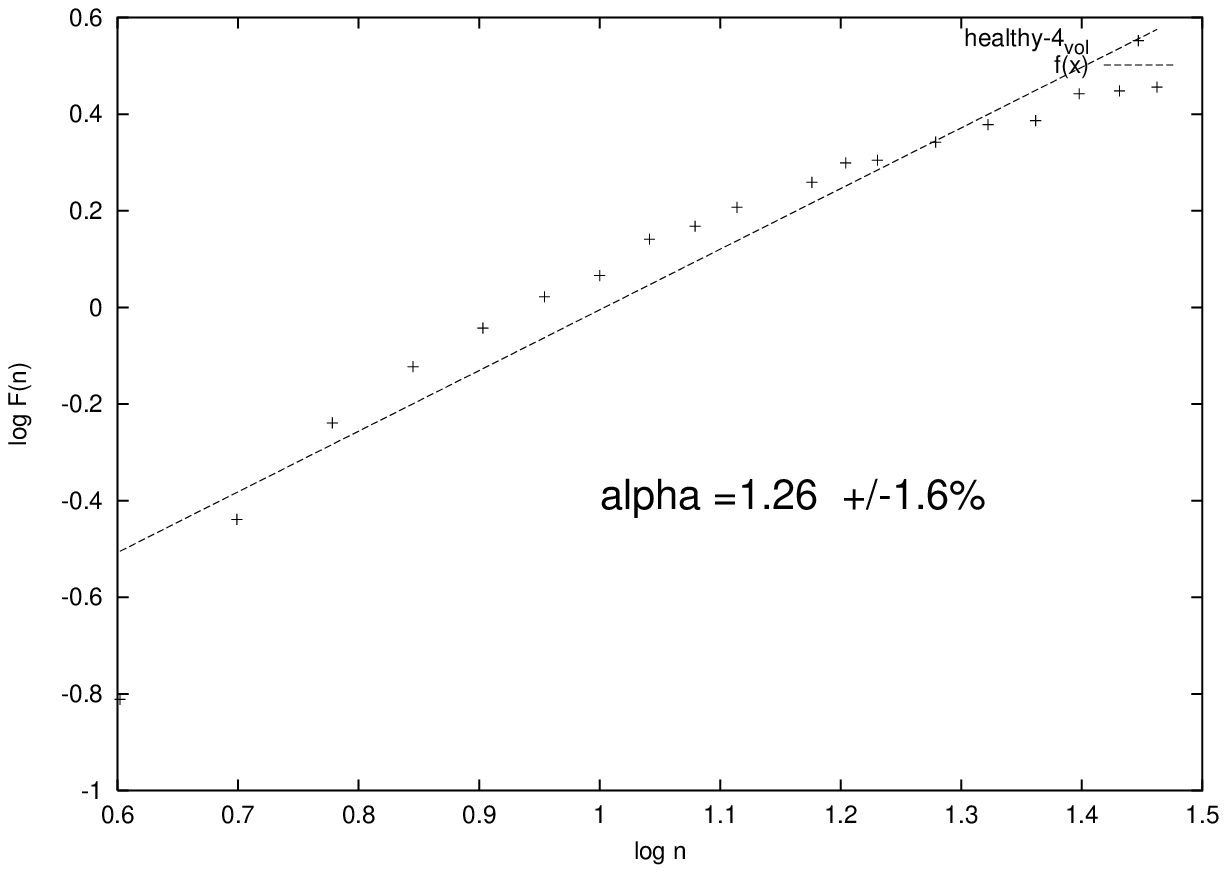}}
\caption{\scriptsize Healthy behaviours -- 
DFA of gallbladder volumes evolution with linear 
fit giving values of  corresponding correlation 
parameter $\alpha$.}
\label{fig:h}
\end{figure}
{
\begin{table}[htbp]
\begin{center}
\begin{tabular}{||l||c|c|||l||c|c||}

\hline
\hline

\textbf{ Healthy} 	&\quad $\alpha$ \quad	& \quad $\pm \%  $&
\textbf{Ill}   	&\quad $\alpha$ \quad	& \quad $\pm \%  $ \\
\hline \hline
	H 1 	& 1.17	& 1.1		&	I 1	&	0.92	&	1.2	\\

	H 2	&	1.40	& 0.5		&I 2 &	1.87	&	0.5	\\

	H 3	&	1.10	&	1.0	&I 3	&	1.03	&	2.2		\\

	H 4	&	1.26	&	1.6		&	I 4	&	1.35	&	0.7	\\

\hline \hline




	
	
\end{tabular}

\caption{\scriptsize{Computed values of the correlation 
factor $\alpha$ regarding gallbladder volume 
 for healthy (left) and ill (right) cases with 
 percentual error. Data set consisting of roughly 
 40 measurements.}}
\end{center}
\label{tab:alpha}
\end{table}
	}
Our data has been taken through 
eco-graphic gallbladder 
volume evaluations of people who was known 
to be ill or healthy (illness related to the 
presence of small balls of fat in the gallbladder, 
viewable in the ecography, thus affecting 
the secretion outflow).
 From our results performed via the 
 Physionet Software for DFA \cite{PHY},
  though preliminary and for 
a small set of data, we see that even if there
is not a sharp distinction between healthy 
and ill situations, some of the ill patients
present an  $\alpha$ with a higher deviation 
from the value 
of unity, showing how the vibration of 
an obstructed  
gallbladder during digestion (for 40 minutes 
after a lunch) can be characterized even by 
not-power-law correlation.

\begin{figure}
\resizebox{\hsize}{0.15\vsize}{\includegraphics{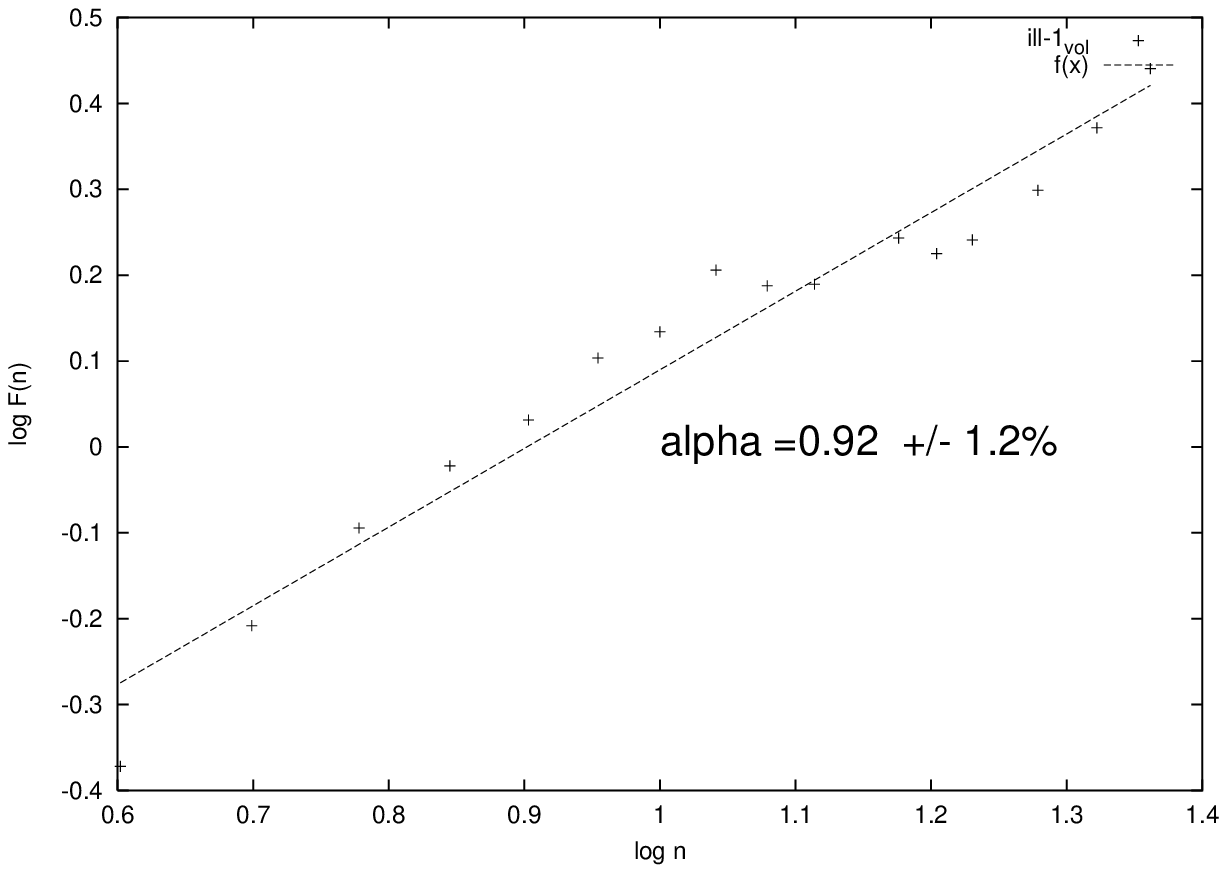}
									\includegraphics{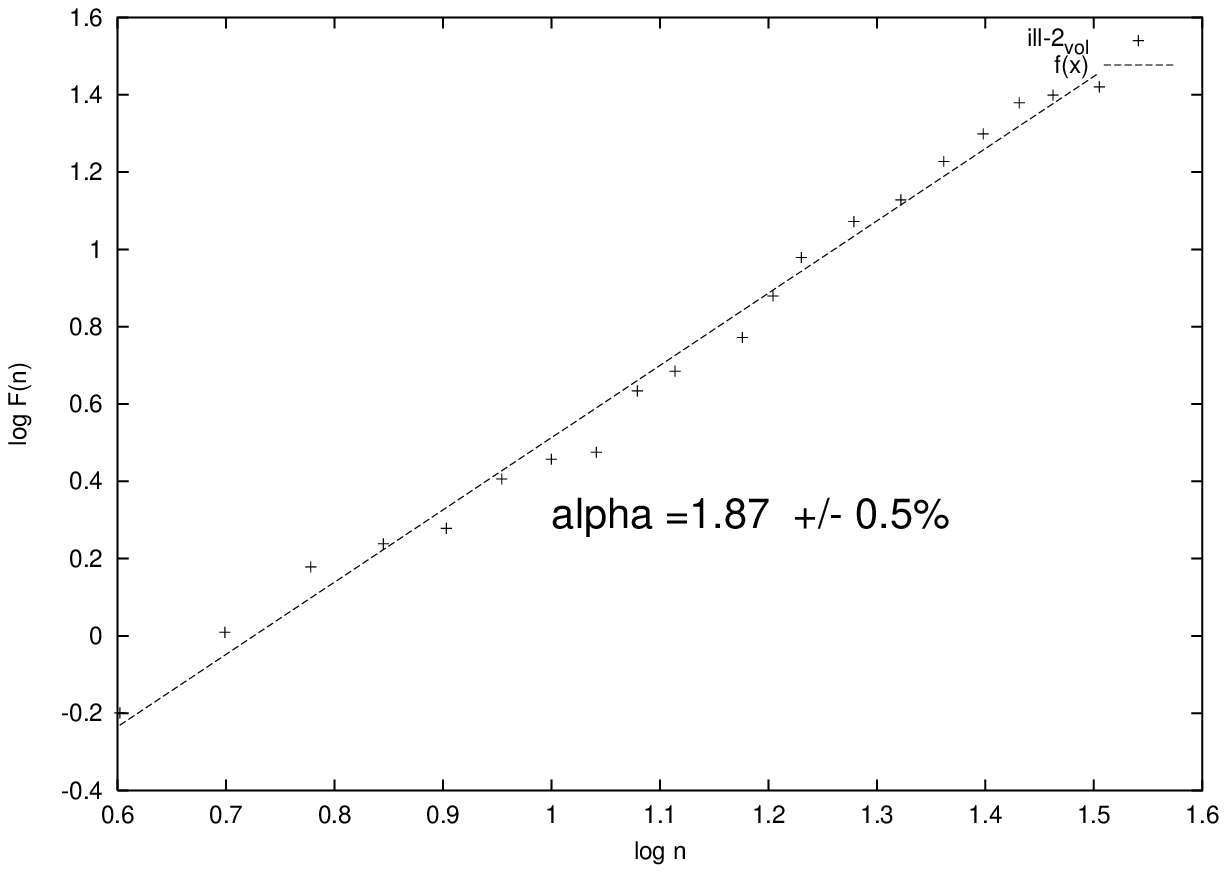}}
\resizebox{\hsize}{0.15\vsize}{\includegraphics{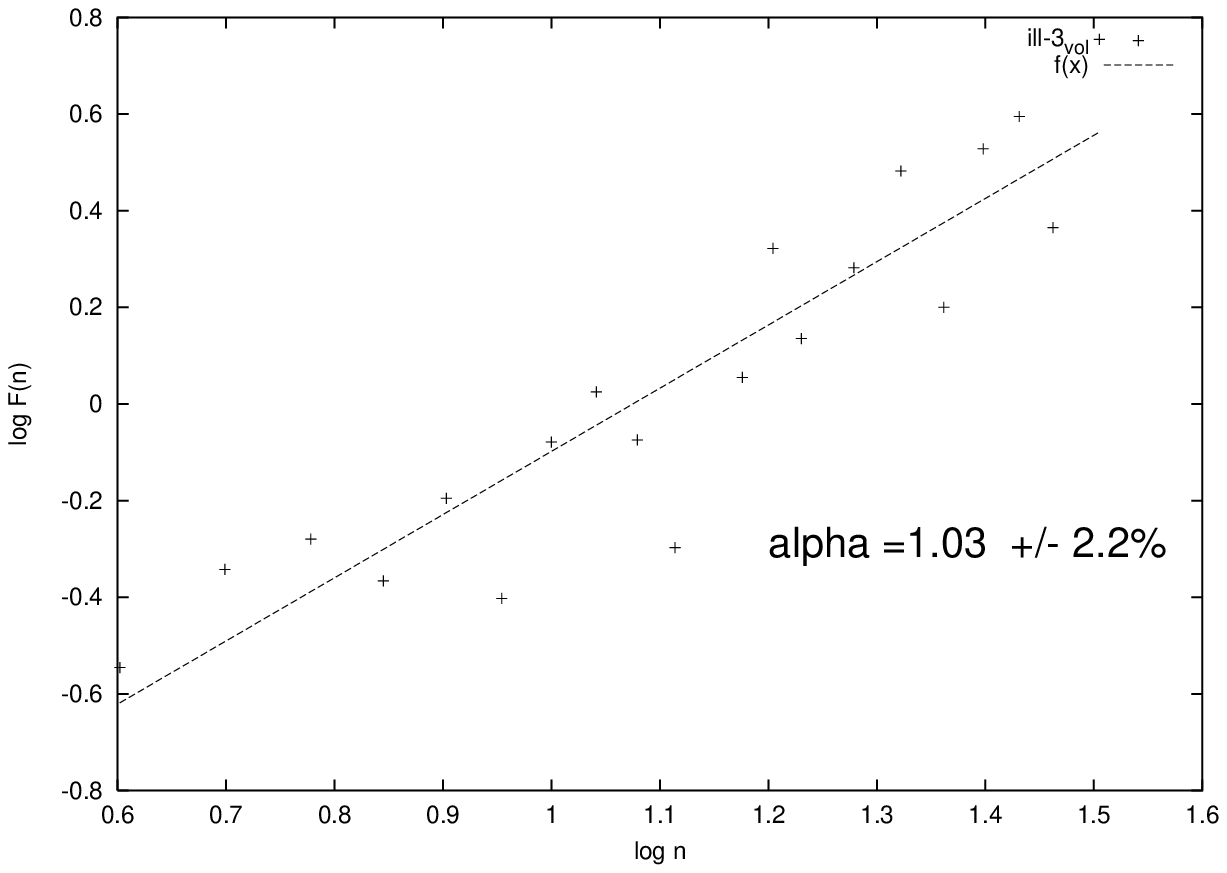}
									\includegraphics{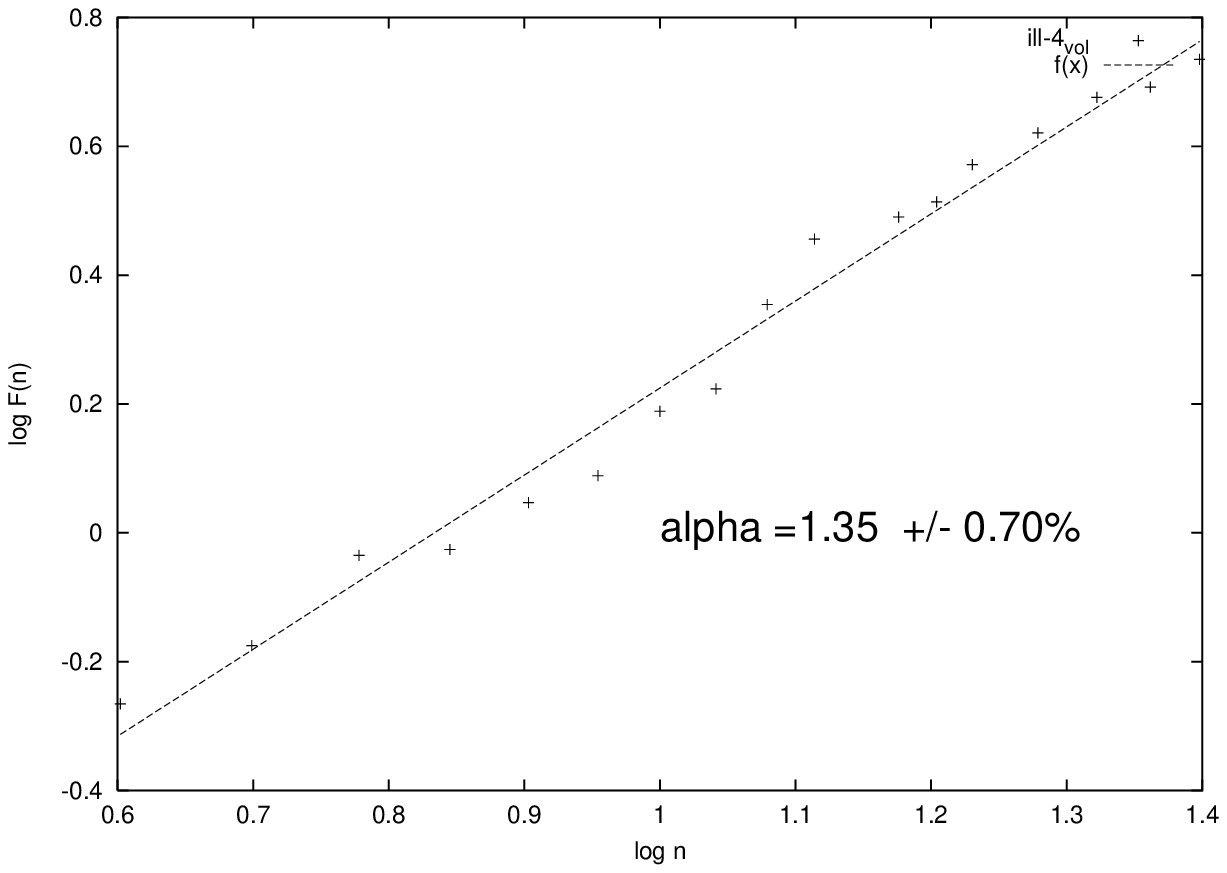}}
\caption{ \scriptsize Ill behaviours -- 
DFA of gallbladder volumes evolution with linear 
fit giving values of  corresponding correlation 
parameter $\alpha$.}
\label{fig:i}
\end{figure}



I would acknowledge Simonetta Filippi for the helpful
discussions and Michele Guarino for providing 
the first raw data.

\end{document}